# Ethical Exploration and the Role of Planetary Protection in Disrupting Colonial Practices

*A submission to the Planetary Science and Astrobiology Decadal Survey 2023-2032*

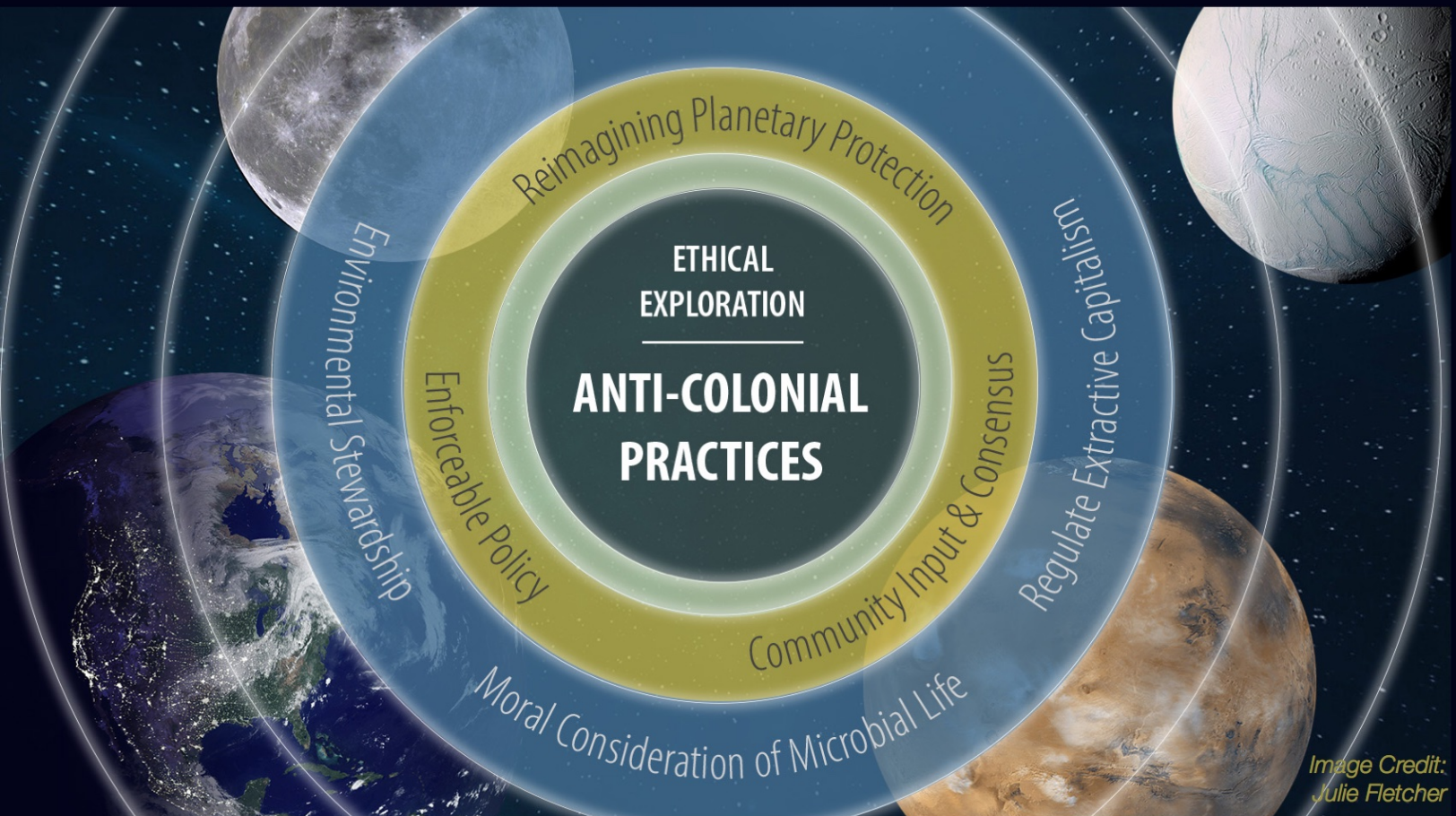

Image Credit: Julie Fletcher

## Frank Tavares

NASA Ames Research Center, frank.j.tavares@nasa.gov, 650-537-3922

**Coauthors:** Denise Buckner (*Blue Marble Space/University of Florida*); Dana Burton (*George Washington University*); Jordan McKaig (*Georgia Institute of Technology*); Parvathy Prem (*Johns Hopkins Applied Physics Laboratory*); Eleni Ravanis (*European Space Agency/ University of Hawaiʻi at Mānoa*); Natalie Trevino (*Western University*); Aparna Venkatesan (*University of San Francisco*); Steven D. Vance (*JPL-CalTech*); Monica Vidaurri (*Howard University/NASA Goddard Space Flight Center*); Lucianne Walkowicz (*The Adler Planetarium/ The JustSpace Alliance*); Mary Beth Wilhelm (*NASA Ames Research Center*)

*Click here for the list of signatories.*

Summary

All of humanity is a stakeholder in how we, the planetary science and astrobiology community, engage with other worlds. Violent colonial practices and structures--genocide, land appropriation, resource extraction, environmental devastation, and more--have governed exploration of Earth, and if not actively dismantled, will define the methodologies and mindset we carry forward into space exploration. With sample return missions from Mars underway, resource maps of the Moon being produced, and private industry progressing toward human exploration of Mars, the timeline is urgent to develop a modern, inclusive, robust, and enforceable policy framework to govern humanity's engagement with other worlds. This paper does not recommend specific policies or implementation strategies, and instead focuses on "the identification of planetary protection considerations," in accordance with the scope of this decadal survey.[1]

Ethical considerations must be prioritized in the formation of planetary protection policy. **The choices we make in the next decade of space exploration will dictate the future of humanity's presence on other worlds, with the potential to impact the environments we interact with on timescales longer than the human species has existed.** We should make these choices consciously and carefully, as many will be irreversible, especially those pertaining to how we interact with potential extraterrestrial life.

It is critical that ethics and anticolonial practices are a central consideration of planetary protection. We must actively work to prevent capitalist extraction on other worlds, respect and preserve their environmental systems, and acknowledge the sovereignty and interconnectivity of all life. The urgency of finding a second home on Mars in the shadow of looming environmental catastrophe on Earth is not only a questionable endeavor[2] but scientifically impossible with present technology,[3] and is often used as a justification for human exploration and to suggest that these ethical questions may be antiquated in the face of that reality. Here we argue the opposite: that the future of our own species and our ability to explore space depends on pursuing anticolonial practices on Earth and beyond. An anticolonial perspective can push us towards an ethic that acknowledges our interconnected and entangled lives. Rather than an escape, or a continuation of manifest destiny, the Moon and Mars may provide the key to practicing other ways of exploring and of being.

**Our primary recommendation is that the planetary science decadal survey call on all bodies involved in developing planetary protection policies to engage in a process of community input to establish a robust reevaluation of the ethics of future crewed and uncrewed missions to the Moon, Mars, and other planetary bodies with the intention of developing anticolonial practices,** centered around the considerations we present. Current policy does not adequately address questions related to in-situ resource utilization, environmental preservation, and interactions with potential extraterrestrial life. Documents such as the Outer Space Treaty in theory address some of these concerns, but lack effectiveness without enforcement mechanisms. A reformed planetary policy that addresses today's challenges and is inclusive, robust, and enforceable must be established. Special emphasis should be placed on resisting colonial structures in these policies.

Our intention is that these considerations become normalized within our field, and engaging with a range of communities and disciplines becomes an automatic part of developing planetary protection policy and in meeting planetary protection requirements during mission design. Until that point, **our second recommendation is that the planetary science and astrobiology community solicits planetary protection input through processes that are interdisciplinary, cross-cultural, and include a multiplicity of epistemologies,** such as but not limited to historians, ethicists, philosophers, and social scientists, as well as the general public. All of these perspectives are essential for creating a responsible policy framework that governs how scientists, engineers, and others interact with other worlds. This expertise is necessary to establish just norms for future space societies. Here we briefly present examples of how space exploration will replicate colonial structures if they are left unchallenged, and the role planetary protection will play in propagating these structures if left unexamined. We also delve into an exploration of the ethical questions that must be addressed on the above topics, including:

- Agency and moral consideration of extraterrestrial microbial life;
- Responsibility towards the potential for future life on any planetary body;
- Parameters for inevitable interactions between microbiomes from different planetary bodies, if humans are to travel to potentially habitable worlds;
- Preservation of environments on planetary bodies for reasons beyond scientific justifications, including historical, environmental, intrinsic, or aesthetic value;
- Long-term environmental impacts of resource extraction on planetary bodies;
- Short-term impact of largely unrestrained resource extraction on wealth inequality.

Colonial Structures Past and Present

Colonial expansion and the trans-Atlantic slave trade have been foundational to our present world. What we call globalization "is the culmination of a process that began with the constitution of America and the colonial/modern Eurocentered capitalism as a new global power."[4] The result is a world where political and economic systems, namely capitalism, prioritize profit over human welfare, producing an environmental crisis[5] and vast inequalities further compounded by climate change.[6] As Roxanne Dunbar-Ortiz writes, "[c]hoices were made that forged that path toward destruction of life itself--the moment in which we now live and die as our planet shrivels, over-heated."[7] Coloniality, the enduring system of domination born from colonialism that we are left with today,[8] is the product of those choices. Understanding what those choices were, and how we are on the precipice of making them again, is essential to ensuring an ethical, anticolonial framework for exploring space.

Several of these mechanisms of colonial violence are of particular relevance, as they connect to current practices in space exploration, planetary science, and astrobiology today:

*Biological Contamination and Ecological Devastation:* The spread of deadly pathogens was used as a form of biological warfare, playing a part in genocide against Indigenous peoples, both intentionally and unwittingly.[9] Colonial expansion caused the population in the Americas to decrease by 90 percent, an enormous loss of life that the dispersal of these pathogens contributed to, alongside concerted warfare and the appropriation of land.[10] All of these practices, including biological contamination, had an impact on the ecosystems Indigenous peoples were a part of--and the destruction of those environments provided an additional way to attack the livelihoods of Indigenous peoples. Settler colonial dominance can be described "as violence that disrupts human relationships with the environment,"[11] a framework that allows us to clearly see how coloniality continues to enact violence on Indigenous lives as well as many other communities through pollution[12] and other environmentally-related effects.

Biological contamination is not a politically neutral or accidental phenomenon and will always have an effect in the environment in which it is taking place amongst all actors involved – both human and nonhuman. This is true for both forward and backward contamination in missions to other planetary bodies. Forward contamination will irreversibly change any extant extraterrestrial microbiome. In the unlikely, but potentially disastrous scenario of backwards contamination, we must also reflect on how structural racism allowed the COVID-19 pandemic to disproportionately impact Black and Indigenous communities.[13] **It is crucial that the planetary science community, with community input, take the opportunity before uncrewed and crewed exploration of other worlds to think ecologically – and seek to equitably address the consequences of our presence on these other worlds.**

*Race Science:* Western science built the lie of racial difference that became a core justification underlying colonial expansion, the slave trade, and genocide against Indigenous peoples.[14] White supremacy is a key aspect of almost all other forms of colonial violence and race science is fundamental to its logic to this day.[15] Moreover, the spread of deadly pathogens and ecological devastation due to colonialism were enabled by this codified disregard for the livelihoods of those that took the brunt of that violence. Modern science is not immune from these antiquated notions-- and science is often invoked to justify and extend racist power structures.[16] There is even peer-reviewed literature replicating the arguments of eugenics--an ideology fundamentally tied to white supremacy in its pursuit of maximizing "favorable characteristics"--in discussions of how reproduction should be handled in future communities on Mars.[17,18] Though the context of human space exploration is a different context with differing moral stakes, white supremacy and related systems of power, if left unchecked, can have deep impacts on future missions. Who we initially send to Mars, and how we come to that decision, will define the nature of those communities going forward.

*Commodification and Appropriation of Land and Resource Extraction:* The commodification of land through extractive practices has led to significant   disruption of the ecosystems that Indigenous communities rely upon for their livelihoods. Examples of extractive exploitation and colonialism abound; while many people in the US think only of the gold rush, mining of rare minerals in Central and South America and Africa incentivize and continue to accelerate colonial expansion even today. Agricultural practices throughout the colonial world have been and continue to be damaging, transforming environments and destroying human lives and cultures.[19] From cotton fields in the American south to sugar plantations and rubber tappers in Brazil, the combination of land and people as property was key to the generation of wealth that built up the Western world.[20]

    **The field of planetary science and space exploration in the present day is not divorced from these practices**, and both existing and planned space infrastructure continue to encroach upon Indigenous land. This is often justified by falsely framing opposition to such encroachments as "obstructions" to "the future."[21] For example, construction of the Thirty Meter Telescope atop Mauna Kea has begun despite opposition from many Kanaka ʻŌiwi (Native Hawaiians), who note that previous astronomy development atop Mauna Kea has already had substantial adverse effects.[22]

    Current structures for in-situ resource utilization on other worlds are analogous to some of these past and current practices on Earth. Most immediately, lunar resource maps seek to enable public and private sector mining actors to plan for extraction of water ice and other resources. Similar proposals exist for asteroid mining. This is presented under a guise of "sustainability," but in actuality replicates the practices of extractive capitalism that have contributed to the environmental degradation of Earth. In the long-term, this exploitative approach to extraterrestrial exploration will be similarly detrimental, and recommendations provided in the white paper "Asteroid Resource Utilization: Ethical Concerns and Progress" address these issues in more depth.[23]

*Public-Private Partnerships as a Colonial Structure:* Private individuals and institutions, in collaboration with governments, are a key aspect of the colonial structure. For example, the East India Company was fundamental to British expansion across the Eastern hemisphere and took a central role in colonial domination and political control as well as trade.[24] More recent examples include the influence of American fruit companies in the United States' interventions into Latin American politics during the Cold War.[25] In the United States, treaties signed with Native American nations have repeatedly been broken, often by settler colonialist individuals working in tandem with the US government and military. The Dakota Access Pipeline, a modern reframing of the ongoing Indigenous demand to honor the Black Hills Treaty,[26] illustrates how capitalist interest intersects with colonialism today.

These examples are mirrored in the active role private industry is currently taking in space exploration. Presently, there is little to no oversight by national governments or international structures. Private partnerships are encouraged to plan missions to the Moon and Mars, often supported by state funding. However, there is a lack of concrete and effective policy to guide their actions, and no consequences are levied when existing policies are violated.[27] For example, the privately-funded and state-operated Beresheet lunar lander crashed on the Moon and accidentally released thousands of tardigrades.[28] At present, bodies like the Moon and Mars are in practice free reign for private entities. An unfortunately accurate euphemism is that we are in a "wild west" of space policy in this regard. When faced with complex and nuanced ethical questions like the ones we will face in space exploration, private actors, by their very structure, will prioritize economic considerations above moral ones. History, through the examples above and others, shows us that they will.

These four points are a vast oversimplification of the tactics used in the multiple centuries of colonial expansion and rule, and do not cover all the ways colonial structures manifest themselves in our field. Instead, we use them to highlight structures pertinent to the ethical issues that planetary protection must tackle in the coming years.

<u>Ethical Questions in Space Exploration</u>

In this paper, we use the term ethics to start a conversation about the moral considerations around our actions in space exploration, recognizing that the term is not well defined within the planetary science and space exploration community. Anticolonial practices are not central to how our field considers ethics, which is why we propose several topics here of immediate concern for the next decade, which we recommend become central considerations in planetary protection going forward.

*Moral Consideration of Extraterrestrial Microbial Life:* There must be further discussion of what moral consideration microbial life on other worlds should have, beyond their scientific significance, as others have considered previously.[29] Considerations of "intelligence" or "non-intelligence" should not be used as the framework for this discussion. Not only do biological distinctions of intelligence have a racist history, they do not hold scientific merit. It is clear that microbiology is foundational to Earth as we know it, and microbes are deserving of moral consideration. We should afford any potential non-terrestrial microbiology on planets like Mars and Venus, or icy moons like Enceladus, Europa, and Titan an even greater consideration, recognizing that extraterrestrial life may also operate in ways not initially obvious to us. What we determine as "intelligence" based on Earth life may not be relevant on another world, compelling us to proceed with caution.

Preliminary models for atmospheric dispersion indicate that even the slightest degree of biological contaminants at any location on the Martian surface will be distributed globally, even when using the most conservative parameters.[30] A human presence on Mars will bring biocontaminants and irreversibly contaminate the planet, both with whole organisms and their chemical constituents. This poses extreme concerns for the ability to conduct sound astrobiology to identify ancient or present life, but a larger moral concern as well. If microbial life currently exists on Mars, any crewed mission may impact that life in ways we cannot predict. **Therefore, it is of paramount importance to consider the ethics of any crewed mission to Mars prior to such an expedition, including an assessment of the structures supporting the project and their intent, to ensure mission design can be impacted by these considerations.**

*Obligations to Potential Future Life:* Even if there is no extant microbial life on Mars or beyond, we must consider the impacts of our actions on geologic timescales. A human presence on an astrobiologically significant world could disrupt evolutionary processes already in place. What moral obligation do we have towards potential future life that our presence on Mars could impact, or to hybrid forms of life that our presence could potentially create? These questions must be addressed by planetary protection policy.

*Ethical Interactions with Potential Microbial Life:* Any first contact scenario with extraterrestrial microbial life will occur at the microbial scale, one that human explorers will not be aware of. It will be a conversation between two microbiomes we will not be privy to and which we will have minimal ability to influence. In the same way a viral or bacterial infection has no understanding of the impacts it causes on a human scale, we will have no understanding of what potential harm our actions are causing.

However, we know enough to prepare for that interaction-- and discuss whether it should occur at all. We must first reject the idea that microbial life is beyond moral consideration due to the label of "non-intelligence" or the claim that Mars is an empty place. We cannot repeat the notions of "terra nullius"[31] that perpetuated colonial violence on Earth. Instead, we must explore anticolonial perspectives and implement those philosophies into our mission designs and scientific practices, letting these guide our approach to interactions with extraterrestrial life. **Our path forward must be an interdisciplinary approach to exploring more thoughtful forms of interaction between these differing microbiomes, with an explicit effort to reject colonial philosophies and structures.**

*Preservation of Environments on Non-Habitable Worlds:* Current plans for the Moon place in-situ resource utilization as a fundamental component of a long-term presence. Current policy does not adequately address questions relevant to preservation beyond sites of scientific value, and ignores questions of whether certain environments should be preserved for historical or environmental reasons, or even their intrinsic value. Aesthetics should also be considered. If Moon mining is to be an extensive enterprise as is planned, those changes will be visible from Earth, fundamentally changing one of the few communal human experiences of gazing at the Moon. In addition, the Moon and other planetary bodies are sacred to some cultures. Is it possible for those beliefs to be respected if we engage in resource utilization on those worlds? Lunar exploration must be prepared to adjust its practices and plans if the answer is no. An alternative approach to how we interact with these environments can be found in Indigenous knowledge, which is inherently interdisciplinary, multigenerational, and expressed through sustainable practices. "Space and place" is an important aspect of Indigenous knowledge, where learning takes place in harmony with a particular place and time. Science in such a framework is not something done "on" or "to" land, but is created in relationship to a place and with deep intentionality and respect.

*Resource Extraction:* Many people consider it inevitable that resource extraction will be a fundamental part of space exploration, if not the reason for doing it at all. However, it is worth questioning whether our current mode of extractive capitalism is something we should take with us when interacting with other worlds. As we are finding on our own Earth, resources are not infinite, and if resource extraction is a cornerstone of the structures we build on the Moon and other worlds, we are setting ourselves up for the same challenges in the long-term. Enabling those with the wealth to privately engage in space exploration efforts could exacerbate already extreme wealth inequality in the immediate future. There are possibilities for engaging in a truly sustainable way with other worlds, and even redistributing resources in an equitable way that doesn't exacerbate wealth inequality. To explore these alternatives and enact them will require conscious consideration when designing policy and robust enforcement mechanisms for their implementation.

Planetary Protection Policy

Space policy as it stands is insufficient to grapple with the questions presented above, and does not have effective enforcement mechanisms in place. The Outer Space Treaty (OST), which on the surface presents as an anticolonial, antimilitarized vision of space that prioritizes preservation over the appropriation of resources and properties, is non-binding, outdated, and does not speak to the realities of today's space industry.

Even recent planetary protection policies, such as those outlined by COSPAR, do not address the moral and ethical rationale for contamination mitigation or other aspects of harmful interference or alteration of extraterrestrial celestial bodies. Furthermore, in the United States there is currently no mechanism to regulate private sector post-launch activities, and thus there is no enforcement mechanism for NASA's planetary protection policies or the values of the OST.

The most recent NASA Interim Directives, [NID 8715.128](#) and [NID 8715.129](#), detailing the planetary protection aspects of the Artemis program, do not address the questions we raise and upon expiration, could lead to even weaker policies than what we have now. On the Moon, only some regions of scientific importance are protected, with the rest open for resource extraction. For Mars, policy is even more loosely defined and fails to consider the global dispersal of terrestrial contaminants. Rather than considering how to enable human exploration to Mars within the confines of past planetary protection policy, those restrictions have been relaxed to enable easier access.

Both of these policies weaken what was previously in place, and give extreme latitude to private industry. Leaving policy structures as they stand will not allow thoughtful consideration of any of the above questions, and will leave these deeply nuanced and complex decisions to unaccountable governments and private companies. Our own history gives us enough data to assume moral considerations will not be prioritized. **Discussions on the ethics of planetary protection must result in robust, enforceable policy that explicitly works to dismantle colonial structures and provide answers or frameworks to address the ethical questions outlined above.** We intentionally did not recommend a separate board or committee to address these issues. These considerations should not be sectioned off as secondary aspects of planetary protection policy, but deemed within the scope of any current or future entities that are developing or advocating for such policy.

Conclusion

    **The space science and space exploration community must address the above concerns and build an enforceable policy structure that seeks to actively dismantle the current systems constructed by the violent past of exploration on Earth.** This must be done not simply to right past wrongs, or ensure ethical interactions with extraterrestrial environments and potential forms of life, but for an ethical and livable future for humanity. The structures we bring to other worlds will ultimately be those that humans live in as well, and in this sense, we need look no further than the existential threats we face here on Earth to see the future: the destruction of our planet's habitability, the ravaging of its ecosystems with disease, and the persistence of racism and other forms of inequity around the globe. Replicating a violent colonial framework will hurt the humans living off-world, retaining our current social inequities and hierarchies and reinforcing those systems on Earth. Ultimately, we must build a better, moral, and livable future because it is how we will survive on our own planet or any other. Proving that we can interact with other worlds in ways that don't reproduce capitalist extraction, that respect and preserve environmental systems, and that acknowledge the sovereignty and interconnectivity of all life, will illustrate these practices are not only possible, but necessary and liberating.

    Dismantling the structures that govern our current world and building new ones will not be easy. We are calling on the decadal committee to engage in that fight, even knowing there will be resistance. Policy is an essential tool in this struggle, but it will require a wider change in philosophy. Space exploration, instead of being the "final frontier," can be a catalyst for a transformative change in how we consider our relationships to other forms of life, to land, and ultimately to each other.

**References**: **[1]** NASEM 2020, Planetary Science and Astrobiology Decadal Survey 2023-2032 **[2]** Walkowicz 2018, The Problem with Terraforming Mars **[3]** Jackosky et al 2018, Inventory of CO2 available for terraforming Mars **[4]** Quijano 2000, Coloniality of Power, Eurocentrism, and Latin America **[5]** IPCC, Climate Change 2014: Synthesis Report **[6]** Diffenbaugh et al 2019, Global Warming has increased global economic inequality **[7]** Dunbar-Ortiz 2014, An Indigenous Peoples' History of the United States **[8]** Quijano 2007, Coloniality and Modernity/Rationality **[9]** Fenn 2010, Biological Warfare in Eighteenth-Century North America: Beyond Jeffrey Amherst **[10]** Dunbar-Ortiz 2014, An Indigenous Peoples' History of the United States **[11]** Whyte 2018, Settler Colonialism, Ecology, and Environmental Injustice **[12]** Fernandez-Llamazares et al. 2020, A State-of-the-Art Review of Indigenous Peoples and Environmental Pollution **[13]** CDC 2020, Health Equity Consideration and Racial and Ethnic Minority Groups **[14]** Saini 2019, Superior: The Return of Race Science **[15]** Painter 2010, The History of White People **[16]** Washington 2006, Medical Apartheid: The Dark History of Medical Experimentation on Black Americans from Colonial Times to the Present **[17]** Szocik et al 2018, Biological and social challenges of human reproduction in a long-term Mars base **[18]** Schuster et al 2016, Mars ain't the kind of place to raise your kid: ethical implications of pregnancy on missions to colonize other planets **[19]** Hippert 2018, Agriculture and Colonialism **[20]** Bhandar 2018, Colonial Lives of Property: Law, Land, and Racial Regimes of Ownership **[21]** Goodyear-Kaʻōpua 2017, Protectors of the Future, Not Protestors of the Past **[22]** Kahanamoku et al. (2020): National Academy of Science Astro2020 Decadal Review: Maunakea Perspectives. figshare. Collection. **[23]** Rivkin et al 2020, Asteroid Resource Utilization: Ethical Concerns and Progress **[24]** Bowen 2009, The Business of Empire: The East India Company and Imperial Britain **[25]** Chapman 2009, Bananas: How the United Fruit Company Shaped the World **[26]** Meyer 2016, *The Legal Case for Blocking the Dakota Access Pipeline* **[27]** O'Shea 2020, The Wild, Wild West of Space Law **[28]** Johnson 2019, The curious case of the transgressing tardigrades (part 1) **[29]** Persson 2012, The moral status of extraterrestrial life **[30]** Piasecny 2019, Atmospheric Dispersal of Contamination Sourced from a Putative Human Habitat on Mars **[31]** Buchan et al 2006, Savagery and Civilization: From Terra Nullius to the 'Tide of History'